\documentclass[aps,pra,reprint,longbibliography,amsmath,amssymb]{revtex4-2}
\usepackage{amsthm,mathtools,graphicx,booktabs,microtype}
\usepackage{xcolor}
\usepackage[colorlinks=true,linkcolor=blue!55!black,citecolor=blue!55!black,urlcolor=blue!55!black]{hyperref}
\newcommand{\Mtwo}{M_2}

\newcommand{\ket}[1]{\lvert #1 \rangle}
\newcommand{\bra}[1]{\langle #1 \rvert}

\newcommand{\Lh}{\widehat L}
\newcommand{\Rcal}{\mathcal R}
\newcommand{\Spa}{S_{\mathrm{PA}}^{\mathrm{var}}}
\newcommand{\Wpartial}{\mathcal W_{\partial}}
\newcommand{\Tr}{\operatorname{Tr}}
\newcommand{\Var}{\operatorname{Var}}
\newtheorem{proposition}{Proposition}
\hypersetup{pdftitle={Information leakage and boundary entropy in a perturbed quantum code},pdfauthor={Luis Lozano}}
\begin{document}
\title{Information leakage and boundary entropy in a perturbed quantum code}
\author{Luis Lozano}
\email{lalozanom@tec.mx}
\affiliation{Tecnol\'ogico de Monterrey, Campus Santa Fe, Mexico City, Mexico}
\date{November 27, 2026}
\begin{abstract}
A spin lost from a quantum memory can reveal which state was stored even when its entropy is the same for every pure input. We study a qubit encoded alongside a Bell pair and allow a weak interaction among the three spins. We quantify the leak by measuring how fast the lost spin becomes correlated with a reference qubit as the interaction is turned on. The curvature of their mutual information at zero coupling equals the part of the quadratic stabilizer R\'enyi coefficient that survives removal of all pairwise perturbations. The direction of the lost spin's polarization can reveal the input. Its length alone determines the entropy. We give the exact condition for all pure inputs to have the same quadratic entropy change and a coupling for which the entropies agree at every strength. We then let the lost spin interact with a Heisenberg chain at infinite temperature. The square of the spin propagator gives the fraction of the initial information curvature visible at each site. In a 41-spin calculation, about 82\% of the information curvature at $t=4\hbar/J$ is carried by Pauli operators acting on at least three spins.
\end{abstract}
\maketitle

\section{Introduction}

Suppose a quantum memory loses one spin. Could someone who finds that spin learn which state was stored? Imagine that two equally likely inputs polarize the spin equally strongly in opposite directions. The spin has the same entropy in both cases. Yet a measurement along the polarization axis lets the observer guess the input correctly more than half the time.

This distinction matters in quantum error correction, where the entropy of a region is often easier to calculate than the information it retains. Consider a memory made of three spins. One holds the input qubit. The other two form a Bell pair, with one partner inside the region we keep and one outside. At first, the outside spin tells us nothing about the input. For a pure input the three spins together remain pure, so the lost spin and the two retained spins have the same entropy. Initially, the Bell pair contributes $\ln 2$ to the entropy of the region. This fixed contribution is the simplest example of the area term in models of holographic quantum error correction \cite{ryu2006_holographic_entropy,almheiri2015_bulk_locality_qec,harlow2017_rt_from_qec}.

A weak interaction among the three spins can let some of the input escape. To measure the leak, we first entangle the input with a separate reference qubit and leave the reference untouched. Any correlation that develops between the reference and the outside spin tells us that the spin has learned something about the input. Our main result connects the initial growth of this correlation to a quantity introduced by Cao, Cheng, Karthikeyan, Li, and Preskill (CCKLP) in their study of perturbed codes \cite{cao2026_state_dependent_magic_codes}. They calculated the part of the encoding's stabilizer R\'enyi entropy that remains after removing changes achievable by acting on pairs of spins. For the memory studied here, that coefficient tells us how quickly reference information escapes as the interaction is turned on. We call it a \emph{witness} because a positive value certifies leakage at quadratic order.

CCKLP subtracted the entropy of the recovered input from the boundary entropy and averaged the remainder over inputs with the same spectrum. Here we follow each pure input separately, before choosing a recovery. We find a coupling for which the lost spin can distinguish opposite inputs along every axis, even though its entropy is identical for all pure inputs. We then place that spin in an interacting chain and ask what an observer would have to measure to find the signal again. Early on, measuring individual spins suffices to see most of it. A few exchange times later, most of the signal is visible only in correlations among several spins.

\section{The encoding and the two experiments}\label{sec:model}

Call the input spin $M$ (the matter qubit) and the two members of the Bell pair $a=G_a$ and $b=G_b$. We refer to this pair as the bond. We keep $A=Ma$ and discard $\bar A=b$. The unperturbed encoder is
\begin{equation}
 V_0\ket\psi=\ket\psi_M\ket{\Phi^+}_{ab},\qquad
 \ket{\Phi^+}=\frac{\ket{00}+\ket{11}}{\sqrt2}.
 \label{eq:base-code}
\end{equation}
A single Hermitian $H$ acts on $Mab$ for every input, with $V_\epsilon=e^{i\epsilon H}V_0$. To measure entropy, prepare pure logical inputs with Bloch vector $u$, trace out $b$, and compute $S(A)=-\Tr\rho_A\ln\rho_A$. The two nonzero eigenvalues at $\epsilon=0$ are $1/2,1/2$. Since $b$ is a qubit, the rank of $\rho_A$ is at most two for every encoding in this experiment.

The entropy experiment uses a pure input on each run. For the leakage experiment, entangle the input with a reference qubit $R$ that never interacts with the memory. The resulting state represents the channel and is called its normalized Choi vector
\begin{equation}
 \ket{J_0}=\ket{\Phi^+}_{RM}\ket{\Phi^+}_{ab},\qquad
 \ket{J_\epsilon}=(I_R\otimes e^{i\epsilon H})\ket{J_0}.
\end{equation}
A Clifford circuit acting on a pure stabilizer state can prepare other stabilizer states. The extra resource needed to prepare states outside this set is called magic \cite{cao2023_nontrivial_area,cao2024_gravitational_backreaction_magic}. For an $n$-qubit pure state the second stabilizer R\'enyi entropy is
\begin{equation}
 \Mtwo(\psi)=-\log_2\!\left(2^{-n}\sum_{P\in\{I,X,Y,Z\}^{\otimes n}}
 \bra\psi P\ket\psi^4\right).
 \label{eq:sre}
\end{equation}
A stabilizer state has precisely $2^n$ Pauli expectations of magnitude one, and all the others vanish \cite{leone2022_stabilizer_renyi}. At $J_0$ the quadratic coefficient of $\Mtwo(J_\epsilon)$ is $Q(H)=4\Var_{J_0}(I_R\otimes H)/\ln 2$ (Appendix~\ref{app:quadratic}). Thus every tangent direction of equal length has the same cost. To find the part that requires all three physical blocks, we remove the changes achievable by acting on pairs.

Let $\mathcal K_\partial$ be the real span of the 36 nonidentity Pauli strings supported on one or two of $M,a,b$. In Eq.~(303) of arXiv version 3 of Ref.~\cite{cao2026_state_dependent_magic_codes}, CCKLP allow an operation on any of the three physical pairs, with $A_1=M$, $A_2=a$, and $\bar A=b$. Their Definition~5.2 removes two-local unitaries to leading order, and their Eq.~(308) expresses the remaining magic as a squared norm with the logical identity component removed. Here we use the index-two entropy in bits and write its minimized quadratic coefficient as
\begin{equation}
 \Wpartial(H)=\min_{K\in\mathcal K_\partial} Q(H-K).
 \label{eq:wpartial}
\end{equation}
Two generators that differ by a phase on $J_0$ give the same tangent to the physical state. The variance removes this phase. Minimizing it projects the centered vector $(H-\langle H\rangle)\ket{J_0}$ orthogonally off the real span of the pairwise vectors. Throughout, operators acting on the physical qubits carry an implicit identity on $R$. Grouping $a$ and $b$ together and allowing $R$ its own block defines the reference-inclusive partition studied in Appendix~\ref{sec:resource-rank}.

\section{What the lost spin can reveal}
\label{sec:mechanism}

Consider any Hermitian $H$ on $MG_aG_b$ and write the logical input as
$\rho(\boldsymbol u)=(I+\boldsymbol u\cdot\boldsymbol\sigma)/2$.
The complement $B=G_b$ starts in $I/2$ for every input.
Let $\boldsymbol r_B(\epsilon,\boldsymbol u)$ be its Bloch vector after
$e^{i\epsilon H}$. In the affine Bloch representation of a qubit channel \cite{ruskai2002}, differentiation gives a vector
$\boldsymbol b$ and a real $3\times3$ matrix $D$ such that
\begin{equation}
  \left.\partial_\epsilon \boldsymbol r_B(\epsilon,\boldsymbol u)
  \right|_0=\boldsymbol b+D\boldsymbol u .
  \label{eq:mechanism-bloch}
\end{equation}
The displacement $\boldsymbol b$ can be measured with the logical input
$I/2$. The three columns of $D$ follow by taking half the difference of
the displacements for the two eigenstates of each logical Pauli. The singular values of $D$ give the semiaxes of the ellipsoid traced by this derivative.
For a qubit channel with Bloch matrix $T$, Wallman \emph{et al.} define a unitarity equal to $\|T\|_{\rm F}^2/3$ \cite{wallman2015}. Here the complementary channel has $T=\epsilon D+O(\epsilon^2)$. The next result identifies this same leading squared norm in CCKLP's projection and compares it with the entropy measured for each input.

\begin{proposition}[Boundary projection and complementary response]
\label{prop:mechanism}
For the product-Bell encoder and the boundary partition $M|G_a|G_b$,
\begin{equation}
  \Wpartial(H)=\frac{\|D\|_{\rm F}^2}{\ln 2}.
  \label{eq:mechanism-witness}
\end{equation}
For pure logical inputs, the entropy of $A=MG_a$ in nats is
\begin{equation}
  S_A(\epsilon,\boldsymbol u)
  =\ln 2-\frac{\epsilon^2}{2}\|\boldsymbol b+D\boldsymbol u\|^2
     + O(\epsilon^3).
  \label{eq:mechanism-entropy}
\end{equation}
Its quadratic coefficient is independent of the pure input if and only if
\begin{equation}
   D^{\mathsf T}\boldsymbol b=0,\qquad
   D^{\mathsf T}D=\lambda I_3\quad\text{for some }\lambda\geq0.
  \label{eq:mechanism-flat}
\end{equation}
\end{proposition}

Appendix~\ref{app:pauli-projection} proves Eq.~\eqref{eq:mechanism-witness} by projecting the Pauli action of $H$ on the two Bell pairs. 
For pure inputs $S(A)=S(B)$, and the two eigenvalues on $B$ are
$(1\pm|\boldsymbol r_B|)/2$. Expanding their entropy at
$\boldsymbol r_B=0$ gives \eqref{eq:mechanism-entropy}, uniformly on the
pure-state sphere. If the squared norm is constant there, its difference
at $\boldsymbol u$ and $-\boldsymbol u$ gives
$\boldsymbol b^{\mathsf T}D\boldsymbol u=0$ for all $\boldsymbol u$.
The remaining quadratic form is constant on the unit sphere precisely
when all eigenvalues of $D^{\mathsf T}D$ agree. The two conditions also
suffice by substitution. The same test applies at any fixed coupling if one uses the exact affine displacement and matrix of the complementary qubit channel. This follows because qubit entropy decreases strictly with Bloch radius on $[0,1]$.

The witness measures the total strength of the input-dependent first
derivative. Entropy measures its length after the common displacement
$\boldsymbol b$ is added. If $D=0$, every input has the same quadratic
entropy coefficient, even when the common displacement is nonzero.
For $D\ne0$, Eq.~\eqref{eq:mechanism-flat} requires
$D=\sqrt{\lambda}\,O$ with $O$ orthogonal and $\boldsymbol b=0$.
The complement then distinguishes directions through its eigenvectors
while its eigenvalues agree to the order governing the quadratic entropy.

An anonymous referee suggested the generator
\begin{equation}
 H_{\rm cyc}=X\otimes Y\otimes Z+Y\otimes Z\otimes X+Z\otimes X\otimes Y.
 \label{eq:mechanism-referee}
\end{equation}
Equation~\eqref{eq:mechanism-contraction} gives
$\boldsymbol b=0$ and $D=2O$ with $O=\operatorname{diag}(-1,1,1)$.
Hence $\Wpartial(H_{\rm cyc})=12/\ln 2$, while
$S_A(\epsilon,\boldsymbol u)=\ln 2-2\epsilon^2+O(\epsilon^3)$ for every pure
input. In this example the direction independence persists at finite
coupling. The three Pauli words anticommute in pairs, so
$e^{i\epsilon H_{\rm cyc}}=c I+i a H_{\rm cyc}$ with
$c=\cos(\sqrt3\,\epsilon)$ and $a=\sin(\sqrt3\,\epsilon)/\sqrt3$.
Writing $\ket{g_j}=(I\otimes\sigma_j)\ket{\Phi^+}$ with $\sigma_0=I$, its action on the encoded pure state is
$c\ket{\psi}\ket{g_0}+a\sum_j O_{jj}\sigma_j\ket{\psi}\ket{g_j}$.
Using
$\Tr_{G_a}\ket{g_i}\bra{g_j}=\sigma_i\sigma_j/2$ yields
\begin{equation}
  \boldsymbol r_B(\epsilon,\boldsymbol u)
  =(2ca-2a^2)O\boldsymbol u .
  \label{eq:mechanism-cyclic-exact}
\end{equation}
Every pure input has the same output Bloch radius at every coupling.

\subsection{How much information reaches the discarded qubit?}

A single entangled input lets us measure $\Wpartial$ without repeating the experiment for different directions. In the Choi experiment, keep the reference $R$ and the discarded qubit $B=b$. They start in the product state $I/4$. Their quantum mutual information measures the total correlation that the encoding has transferred across this cut. In bits we obtain
\begin{equation}
 I(R:B)_{J_\epsilon}=\tfrac12\epsilon^2\Wpartial(H)+O(\epsilon^3).
 \label{eq:leakage}
\end{equation}
To see the normalization, expand the input Bell projector as $(I\otimes I+\sum_i\sigma_i^{\mathsf T}\otimes\sigma_i)/4$. Applying the complementary channel gives $\rho_{RB}=I/4+\epsilon\dot\rho_{RB}+O(\epsilon^2)$, with
\[
 4\dot\rho_{RB}= I\otimes(\boldsymbol b\cdot\boldsymbol\sigma)+
 \sum_i\sigma_i^{\mathsf T}\otimes[(D e_i)\cdot\boldsymbol\sigma],
\]
where $e_i$ are the Cartesian unit vectors. For a traceless perturbation $X$ about $I/d$, the entropy in bits is $\log_2 d-d\epsilon^2\Tr(X^2)/(2\ln 2)+O(\epsilon^3)$. Orthogonality of the Pauli matrices then gives $S(RB)=2-\epsilon^2(|\boldsymbol b|^2+\|D\|_F^2)/(2\ln2)+O(\epsilon^3)$. The displacement term cancels in $S(R)+S(B)-S(RB)$. Since $RAB$ is pure and $S(R)=1$, the coherent information from $R$ to $A$ before decoding is exactly $1-I(R:B)$. Small coherent-information loss allows approximate recovery \cite{schumacher1996_coherent_information,schumacher2002approximate}. Equation~\eqref{eq:leakage} therefore quantifies the initial loss of coherent information through the cut. Appendix~\ref{app:leakage-dimensions} derives the same equality between information curvature and the projected pure-state Fisher metric for arbitrary logical and bond dimensions. 

\section{Two couplings with the same witness}\label{sec:equal}

\begin{figure}[htbp]\centering
\includegraphics[width=\linewidth]{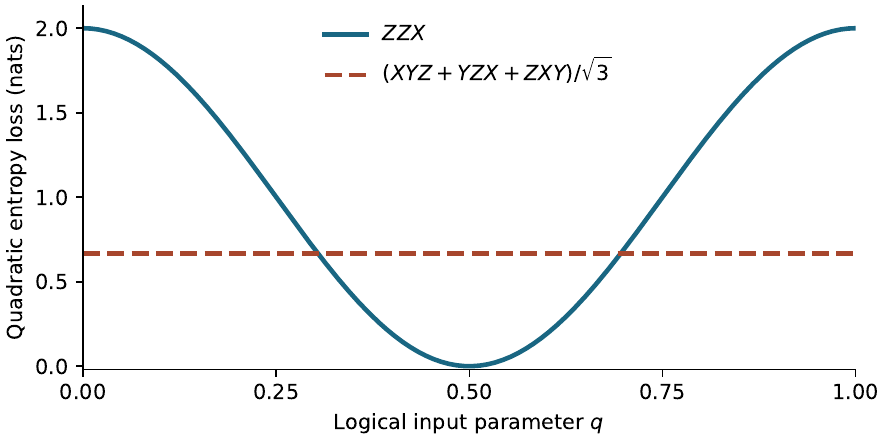}
\caption{Coefficient of $\ln2-S(A)$ at order $\epsilon^2$ for two equal-witness Hamiltonians. At $q=1/2$ the input lies along $X$. The single Pauli coupling then has zero quadratic entropy loss. The cyclic coupling has loss coefficient $2/3$ throughout the meridian.}
\label{fig:matched}
\end{figure}

Consider $Z_M Z_a X_b$ and the normalized cyclic coupling $(X_M Y_a Z_b+Y_M Z_a X_b+Z_M X_a Y_b)/\sqrt3$. Both square to the identity, so their operator norms and coupling scales agree. Each has $\Wpartial=4/\ln2$. For the first, the complement receives one component of the input, with polarization length $2|u_z|\,|\epsilon|$ to first order. For the second, the three components reach the complement with equal strength, and every pure input produces length $2|\epsilon|/\sqrt3$ to the same order. Figure~\ref{fig:matched} compares them on the meridian $\ket{\psi_q}=\cos(\pi q/2)\ket0+\sin(\pi q/2)\ket1$. For the cyclic coupling an observer could distinguish the two poles by measuring a spin component on $b$, even though measuring the eigenvalues at either pole gives the same result.

For a uniformly distributed pure input $u$, let $k(u)=|\boldsymbol b+Du|^2/2$. The mean loss coefficient is $\mathbb E_u k=|\boldsymbol b|^2/2+(\ln 2)\Wpartial/6$. The witness thus fixes the input-dependent contribution to the mean loss. To test how much the loss varies with the input, write $G=D^{\mathsf T}D$. The variance is
\begin{equation}
 \Var_u k = \frac{|D^{\mathsf T}\boldsymbol b|^2}{3}
       +\frac{\|G-(\Tr G)I/3\|_F^2}{30}.
 \label{eq:angular-variance}
\end{equation}
Indeed $\mathbb E[u_i u_j]=\delta_{ij}/3$, and the fourth moment is the sum of the three pairings of Kronecker deltas divided by 15. The term linear in $u$ is odd and has zero covariance with the quadratic term. The first term measures the overlap of the displacement with the input signal. The second measures the spread of the squared semiaxes. This variance tests state dependence across the entire input sphere.

\section{Following the leaked signal through a spin chain}\label{sec:spin-chain}

\begin{figure*}[t]
\centering
\includegraphics[width=.89\textwidth]{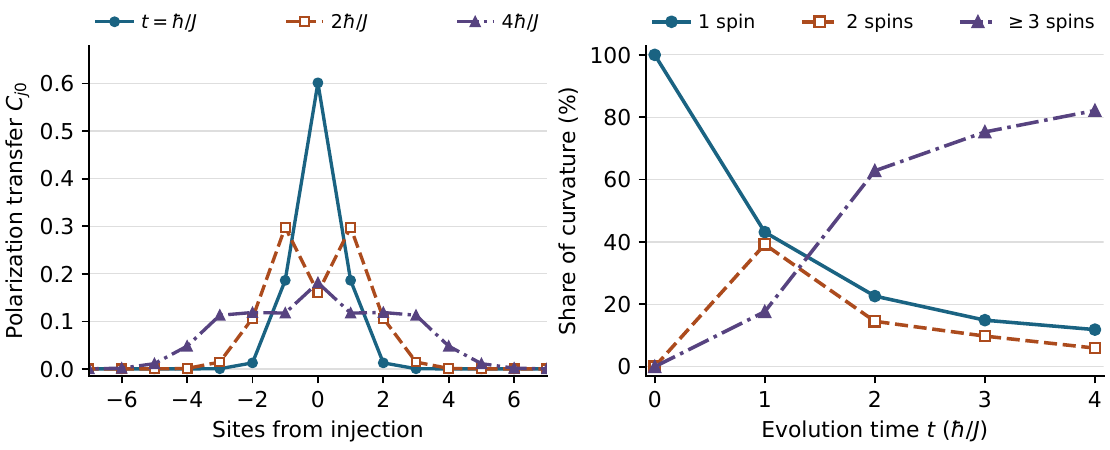}
\caption{Transfer of the injected polarization in a 41-spin open Heisenberg chain. Left, signed spin response at three times (central 15 sites shown). Right, fraction of the initial curvature in observables supported on one, two, or at least three spins. Points give the computed values and lines join them. Time is in units of $\hbar/J$.}
\label{fig:spin-chain}
\end{figure*}

Once a spin has escaped the code, interactions with its surroundings can change where the information is visible. Take the discarded qubit $b$ as the central site, labelled 0, of an open spin-$1/2$ chain. Prepare the other spins independently in $I/2$. After the cyclic injection of Eq.~\eqref{eq:mechanism-referee}, isolate the reference and encoder, and evolve only the chain with
\[
 H_{\rm chain}=J\sum_{j}\boldsymbol S_j\cdot\boldsymbol S_{j+1},
 \qquad \boldsymbol S_j=\boldsymbol\sigma_j/2 .
\]
The injection strength $\epsilon$ and the subsequent evolution time $t$ are independent. The spin response at infinite temperature is well studied for this Hamiltonian \cite{ljubotina2019}. Here it determines how much of the initial reference correlation one can detect at a given site.

Write $U_t=\exp(-it H_{\rm chain}/\hbar)$ and define the transfer of a unit polarization by
\begin{equation}
 C_{j0}(t)=2^{-N}\Tr[Z_j U_t Z_0 U_t^\dagger] .
 \label{eq:spin-propagator}
\end{equation}
The channel from site 0 to site $j$ sends $I/2$ to $I/2$. It also commutes with all spin rotations, since both the dynamics and the bath preparation have this symmetry. Its real Bloch matrix is therefore a scalar multiple of the identity. Inserting $Z_0$ gives the scalar in Eq.~\eqref{eq:spin-propagator}. If the cyclic injector gives the initial spin polarization $s(\epsilon)O u$ of Eq.~\eqref{eq:mechanism-cyclic-exact}, the polarization at site $j$ is exactly $C_{j0}(t)s(\epsilon)O u$. For fixed $j,t,\epsilon$ every pure input yields the same single-site entropy. Opposite input directions can nevertheless be distinguished wherever the polarization is nonzero.

The untouched reference supplies a quantitative test. Let $W_j(t)$ denote the second derivative of $I(R:j)$ in bits with respect to $\epsilon$ at zero. Each of the three input-dependent Pauli components of the Choi derivative is multiplied by $C_{j0}(t)$. Equation~\eqref{eq:leakage} gives
\begin{equation}
 W_j(t)=C_{j0}(t)^2\Wpartial,\qquad
 \left.\partial_\epsilon^2 I(R:\mathrm{whole\ chain})\right|_0=\Wpartial .
 \label{eq:transported-curvature}
\end{equation}
The second identity follows because the added spins start independent and $U_t$ acts unitarily on the chain. For this unnormalized cyclic Hamiltonian, $\Wpartial=12/\ln 2$. The first identity concerns curvature at fixed time. The full mutual information at finite injection strength is nonlinear in the signed transfer, as shown in Appendix~\ref{app:spin-checks}.

To locate the part invisible in one-spin marginals, expand $U_t Z_0 U_t^\dagger$ in products of Pauli matrices normalized with inner product $2^{-N}\Tr(P^\dagger Q)$. The sum of the squared coefficients is one. The identity coefficient vanishes, and rotation symmetry restricts its one-spin coefficients to $C_{j0}$ on $Z_j$. Thus $\sum_j C_{j0}^2$ is the one-spin share of the curvature. The complementary share belongs to products supported on two or more sites. A global quarter-turn converts the injected $Z$ into $X$ or $Y$ and preserves support size, so all three components have the same distribution across these groups. These products can be measured by retaining joint outcomes of local spin measurements. Operator spreading is commonly described by these squared coefficients \cite{khemani2018}. McGinley \emph{et al.} relate them to purities of reduced Choi states \cite{mcginley2022}. Here each squared coefficient contributes to the curvature of the von Neumann mutual information for the weakly perturbed encoding.

In Fig.~\ref{fig:spin-chain}, at $t=4\hbar/J$ about $12\%$ of the curvature is visible in measurements of individual spins. Products on exactly two spins account for a further $6\%$. The remaining $82\%$ comes from products on at least three spins. The left panel shows how far the polarization has traveled in that time. The change in the three fractions on increasing the chain length from 21 to 41 at the same numerical resolution was less than $1.2\times10^{-7}$ in absolute fractional units. 

A two-spin control explains why the joint measurement matters. Coherently applying $\sqrt{\mathrm{SWAP}}$ and randomly swapping the spins with probability $1/2$ give identical individual-spin channels. Each retains curvature $\Wpartial/4$. Measuring the pair distinguishes the two mechanisms. The coherent gate retains $\Wpartial$ jointly and the random channel, with its classical record discarded, retains $\Wpartial/2$. The difference is a correlation present only in the coherent exchange.

\section{What can the memory recover?}\label{sec:example}

Return to the two spins retained in the memory. How well can they recover the stored qubit? As an example, consider a gate that acts only when the input spin and the near end of the bond are both in $\ket1$. Its generator is $H=P_{1,M}P_{1,a}X_b$, where $P_1=(I-Z)/2$. If the matter is $\ket0$, the bond is untouched. If it is $\ket1$, only the $\ket{11}$ branch of the bond rotates. For an arbitrary superposition of these two inputs, tracing out $M$ weights the two branches by their populations. Writing $p=|\langle1|\psi\rangle|^2$, we obtain
\begin{equation}
 S(A)=h\!\left(\frac{1+p\sin\epsilon}{2}\right),
 \label{eq:signal-entropy}
\end{equation}
where $h(x)=-x\ln x-(1-x)\ln(1-x)$. In the rotated branch the bond vector is $(\ket{00}+\cos\epsilon\ket{11}+i\sin\epsilon\ket{10})/\sqrt2$. Its far leg has Bloch vector $(0,-\sin\epsilon\cos\epsilon,\sin^2\epsilon)$, while the untouched branch has zero polarization. The mixture thus has radius $p|\sin\epsilon|$. Over all pure inputs the maximum is at $p=0$, the minimum at $p=1$, and their difference is $\ln2-h((1+\sin\epsilon)/2)=\epsilon^2/2+O(\epsilon^4)$. Appendix~\ref{app:quadratic} works through the subtraction for this gate, giving $\Wpartial=1/(4\ln2)$.

Even keeping $D$ fixed leaves room to change the entropy response. Add $\lambda Z_aX_b$ to the controlled Hamiltonian. This pairwise term drops out of the minimization and shifts the first-order polarization along $Y_b$ by $2\lambda$. The quadratic loss becomes $(2\lambda-p)^2/2$. As $p$ ranges from zero to one, its range falls from $1/2$ at $\lambda=0$ to $1/8$ at $\lambda=1/4$. The two encodings have the same information-leakage curvature in Eq.~\eqref{eq:leakage}, yet their entropy contrasts differ by a factor of four.

Six simpler gates help identify which part of the coupling matters. Rotations generated by $X_M$, $X_a$, or $P_{1,M}X_a$ act entirely inside $A$ and preserve the entanglement across the cut. A controlled $P_{1,M}X_b$ rotation has the same action on the initial Bell pair as $P_{1,M}X_a$, since $X_b\ket{\Phi^+}=X_a\ket{\Phi^+}$. For $X_M X_a X_b$, the product $X_a X_b$ fixes the bond, leaving a matter rotation. Finally $P_{1,a}X_b$ changes the bond entropy to $h((1+\sin\epsilon)/2)$ equally for all matter states. It contributes a common displacement to every output. At $\epsilon=\pi/8$ all seven gates generate positive total $\Mtwo$ on the Choi state. The matter-controlled bond move has the smallest value, $0.214404$ bits.

A decoder must use the same operation for every possible input. We let a two-qubit unitary act on $A$ and discard one output spin. We choose the unitary by searching for high coherent information, which measures how well the remaining output preserves its correlations with the reference \cite{schumacher1996_coherent_information}. A second calculation optimizes the entanglement fidelity over all recovery channels \cite{schumacher1996_entanglement_channels,fletcher2007_optimum_recovery_sdp}.

At $\epsilon=\pi/8$, $A$ retains $0.972383$ bits of coherent information before decoding. The selected decoder recovers $0.921821$ bits and attains Bell fidelity $0.990393$, equal to the numerical fidelity optimum to six decimal places. The Supplemental Material gives an explicit decoder with those values \cite{supplement}. A companion manuscript derives its fidelity certificate \cite{lozano_companion}.

We can now compare the entropy of the retained region with that of the decoded qubit. Their difference is CCKLP's \emph{proto-area}, evaluated here with one fixed decoder. Li studies what further conditions would allow these differences on several cuts to represent areas in one geometry \cite{li2026_nongeometry_witnesses}. For our chosen decoder $\Rcal_U$ the difference at one cut is
\begin{equation}
 \Spa(q,A)=S(\rho_A(q))-S\!\left(\Rcal_U(\rho_A(q))\right).
 \label{eq:spa}
\end{equation}
On the 21 equally spaced points $q=0,0.05,\ldots,1$, the sampled ranges of boundary entropy, recovered entropy, and their difference are $0.075124$, $0.054189$, and $0.084924$ nats, respectively. The two entropies reach their extrema at different inputs. A useful control applies two CNOTs, both controlled by $b$, first targeting $M$ and then $a$. This Clifford circuit gives a boundary entropy range of $\ln 2$ but only $1/2$ Bell fidelity after recovery. A measurement of boundary entropy alone would miss that loss of logical information.

\section{Discussion}\label{sec:discussion}

Suppose the sender prepares one of two opposite pure inputs, with equal probability. An observer on the discarded qubit measures spin along their output polarization difference. The trace distance of the two output states, $T=\tfrac12\|\rho_+-\rho_-\|_1$, gives a probability $(1+T)/2$ of guessing the input correctly. For a pair along a logical axis $u$, Eq.~\eqref{eq:mechanism-bloch} yields $T=|\epsilon|\,|Du|+O(\epsilon^2)$, since the common displacement cancels. With the cyclic coupling the observer has the same advantage along every axis, including when the two output entropies are exactly equal. For the controlled bond rotation the two $X$ eigenstates have identical populations and give identical outputs at every coupling. These measurements distinguish the mechanisms directly. Squaring the three trace distances for the Cartesian input pairs and adding gives $\epsilon^2(\ln2)\Wpartial$ to leading order.

In the chain, the same guessing experiment has a different outcome depending on how many spins the observer keeps. A one-spin measurement reads only the part of the polarization that has reached that site. Keeping two measurement outcomes also gives their product. At $t=4\hbar/J$ even all the one- and two-spin terms together account for only about $18\%$ of the response. The rest can still be found in joint measurements, because evolution of the entire chain preserves its correlation with the reference. This calculation follows the signal for a finite time. Establishing thermal equilibrium would require information about the later dynamics.

The meaning of ``pairwise'' also depends on who holds each spin. If both bond spins belong to a single block $G$, any physical interaction on $MG$ already acts on a pair. Appendix~\ref{sec:resource-rank} gives the more general count when an interaction can also act on the reference. Appendix~\ref{app:two-bond} illustrates the effect of regrouping the spins for two bonds with unequal entropies.

To interpret a boundary entropy as a geometric area, one needs an independent way to measure the area. In Chandrasekaran's null-cone construction the corner area generates a relative boost whose effect can be read from the correlation between two shear probes [Eq.~(5.29) of Ref.~\cite{chandrasekaran2026null}]. A code that aims to reproduce that measurement must identify a boost generator and corresponding probes. Our code answers a question that can be tested with the spins themselves. Preparing three pairs of opposite inputs and measuring the lost spin determines the leading leakage through the cut.

\section*{Computational methods}
The recovery search used four starts and seed 42. For the joint output $\widetilde\omega$ of the reference $R$ and decoded qubit $\Lh$, the objective was $S(\widetilde\omega_{\Lh})-S(\widetilde\omega)$ in bits. A final rotation aligned the decoded basis with the reference Bell state for the fidelity measurement. The fidelity optimization used an input-first, unnormalized Choi matrix with $J\geq0$ and $\Tr_{\Lh}J=I_A$. Appendix~\ref{app:spin-checks} describes the chain evolution, convergence checks, and Lean proofs. AI assistance with derivations and code is described below. We checked the leakage calculation by independent partial traces and Pauli projections, and compared the chain evolution with exact small-system calculations.

\begin{acknowledgments}
The author thanks an anonymous referee for suggesting the cyclic coupling and asking how the pairwise tangent space depends on Schmidt rank.
\end{acknowledgments}

\section*{Data availability}
The Supplemental Material \cite{supplement} gives the recovery data and solver comparisons. The numerical data and code for both figures, the recovery calculations, and the spin-chain convergence tests are available from the corresponding author upon reasonable request.

\section*{Conflicts of Interest}
The author declares no competing interests.

\section*{AI Use Disclosure}
OpenAI Codex assisted with derivations, programming, and drafting. The author proposed the many-spin application and requested formal checks of the analytical claims. Reviews used OpenAI's GPT-6/Astra family, Anthropic Claude Fable~5.1 and 5.2, Kimi CLI, and Google's Gemini~4 Neon and Carbon. Earlier reviews also used Sakana Fugu and Fugu Ultra. The Kimi backend version was not recorded. These reviews suggested tests of rotation symmetry, recovery, and the distinction between local and joint measurements. Appendix~\ref{app:spin-checks} and the Supplemental Material \cite{supplement} describe the direct numerical comparisons and the statements checked in Lean.

\appendix
\section{Projection in the Bell basis}\label{app:pauli-projection}
We prove Eq.~\eqref{eq:mechanism-witness} by evaluating both sides in the same basis.

Set $\sigma_0=I$ and expand
$H=\sum_{i,j,k}h_{ijk}\sigma_i\otimes\sigma_j\otimes\sigma_k$
with real coefficients and indices in $\{0,x,y,z\}$. In the following
sums $j,k$ are spatial indices. Write $\sigma_j^{\mathsf T}=t_j\sigma_j$
with $(t_x,t_y,t_z)=(1,-1,1)$ and let $\varepsilon_{\ell jk}$ be the
Levi--Civita symbol. The Bell expectation on $G_aG_b$ is
$\langle\sigma_j\otimes\sigma_m\rangle=t_j\delta_{jm}$ for spatial $j,m$.
Taking $i\langle[\sigma_{\ell,B},H]\rangle$ in the undeformed state gives
\begin{equation}
 C_{\ell i}=2\sum_{j,k}t_j\varepsilon_{\ell jk}h_{ijk}.
 \label{eq:mechanism-contraction}
\end{equation}
Here $b_\ell=C_{\ell0}$ and $D_{\ell i}=C_{\ell i}$ for $i=x,y,z$.

To compare with the boundary projection, use the orthonormal vectors
$\ket{m_i}=(I\otimes\sigma_i)\ket{\Phi^+}_{RM}$ and
$\ket{g_\ell}=(I\otimes\sigma_\ell)\ket{\Phi^+}_{G_aG_b}$, including
$i,\ell=0$. The physical centered action is
$(I-\ket{J_0}\bra{J_0})(I_R\otimes H)\ket{J_0}$.
We omit the common factor $i$ of unitary tangents throughout the projection.
The one- and two-block actions span the following mutually orthogonal
directions over the reals, with $i,\ell$ spatial:
\begin{equation}
 \ket{m_i g_0},\quad
 \ket{m_0 g_\ell},\quad i\ket{m_0 g_\ell},\quad
 \ket{m_i g_\ell}.
 \label{eq:mechanism-span}
\end{equation}
Indeed, a single matter or bond-leg Pauli gives one of the first two
directions. A product of distinct Pauli axes on the two bond legs gives
the third, and a matter Pauli times a single bond-leg Pauli gives the
fourth. Equal axes on the two legs give only a multiple of the base on
the bond. This exhausts the pairwise generators and gives real rank $18$.

For a three-body Pauli, Bell transposition replaces its bond factors by
$\sigma_k\sigma_j^{\mathsf T}$ acting on $G_b$. Distinct axes therefore
give $-i t_j\sum_\ell\varepsilon_{\ell jk}\sigma_\ell$, whereas equal
axes return to $\ket{g_0}$ with a real coefficient. After the projection
onto the real orthogonal complement of \eqref{eq:mechanism-span}, the
physical centered action is exactly
\begin{equation}
   -\frac{i}{2}\sum_{i,\ell=x,y,z}D_{\ell i}\ket{m_i g_\ell}.
   \label{eq:mechanism-residual}
\end{equation}
Squaring and multiplying by $4/\ln 2$ gives
\eqref{eq:mechanism-witness}.

To compare with CCKLP, denote their operator $D$ in Eq.~(153) of
Ref.~\cite{cao2026_state_dependent_magic_codes} by $K$ and set
$W_R=H$, $\chi=\ket{\Phi^+}\bra{\Phi^+}_{G_aG_b}$.
Bell transposition gives
$K=2i\,\Tr_{G_b}[H,I_M\otimes\chi]
=\sum_{i=0,x,y,z}\sum_{\ell=x,y,z}C_{\ell i}\,
 \sigma_i\otimes\sigma_\ell^{\mathsf T}$ on $MG_a$.
The minimization over the three pairs in their Eq.~(303) leaves the trace
combination in their Eq.~(308),
$\Tr(K^2)-\tfrac12\Tr[(\Tr_M K)^2]=4\|D\|_{\rm F}^2$,
since the partial trace subtracts the $i=0$ displacement.
At $\alpha=2$ and $d=d_\chi=2$ their prefactor is $1/4$, so converting
their natural-logarithm coefficient to bits gives
$\|D\|_{\rm F}^2/\ln2=\Wpartial(H)$.

\section{Complementary leakage for arbitrary register dimensions}
\label{app:leakage-dimensions}

Let the logical register and its physical image $M$ have dimension $d$,
and let each leg of the geometry bond have dimension $g$. The encoder is
$V_0\ket\psi=\ket\psi_M\ket{\Phi_g}_{G_aG_b}$, where
$\ket{\Phi_n}=n^{-1/2}\sum_{j=1}^n\ket{jj}$. The normalized Choi vector
is $\ket{J_0}=\ket{\Phi_d}_{RM}\ket{\Phi_g}_{G_aG_b}$.
For a Hermitian $H$ on $MG_aG_b$, let $W_{\rm F}(H)$ be $4/\ln 2$ times
the squared real distance of its centered action on $\ket{J_0}$ from
the span of the one- and two-block Hermitian actions on $M|G_a|G_b$.
With $B=G_b$, the normalized Choi state of $V_\epsilon=e^{i\epsilon H}V_0$
satisfies
\begin{equation}
 I(R:B)_\epsilon=\frac{\epsilon^2}{2}W_{\rm F}(H)+O(\epsilon^3)
 \qquad \text{(bits)} .
 \label{eq:dimensional-leakage}
\end{equation}
This is a statement about the projected pure-state Fisher metric in any
finite dimensions. When $d$ and $g$ are powers of two, the Bell registers
in the qubit computational frame form a stabilizer state and $W_{\rm F}$
also equals the projected quadratic coefficient of $M_2$.

Choose Hermitian bases $A_\alpha$ on $M$ and $B_\beta$ on $G_b$ with
identity at index zero and $\Tr(A_\alpha A_\gamma)=d\,\delta_{\alpha\gamma}$,
$\Tr(B_\beta B_\nu)=g\,\delta_{\beta\nu}$. Expand
$H=\sum_\alpha A_\alpha\otimes K_\alpha$ with Hermitian $K_\alpha$ on
the bond and define $C_\alpha$ by
\begin{equation}
 K_\alpha\ket{\Phi_g}=(I\otimes C_\alpha)\ket{\Phi_g},\qquad
 C_\alpha=\sum_\beta z_{\alpha\beta}B_\beta .
 \label{eq:dimensional-mirror}
\end{equation}
The coefficient $z_{\alpha0}=\bra{\Phi_g}K_\alpha\ket{\Phi_g}$ is real.
With $\ket{m_\alpha}=(I\otimes A_\alpha)\ket{\Phi_d}$ and
$\ket{g_\beta}=(I\otimes B_\beta)\ket{\Phi_g}$, the $z_{\alpha\beta}$
are the coefficients of the generator action in an orthonormal product
basis. Centering removes $z_{00}$. An arbitrary bond Hermitian supplies
both real quadratures of each $\ket{m_0g_\beta}$ for $\beta>0$.
A matter Hermitian supplies the real $\ket{m_\alpha g_0}$ directions.
For $\alpha,\beta>0$, all real $\ket{m_\alpha g_\beta}$ occur through
$MG_b$ interactions. $MG_a$ interactions give real combinations of
the same directions because transposition preserves Hermiticity.
These exhaust the allowed pairwise supports. Define the real transfer
coefficients and their input-independent part by
\begin{equation}
 D_{\beta\alpha}=-2\,\operatorname{Im} z_{\alpha\beta},
 \qquad b_\beta=-2\,\operatorname{Im}z_{0\beta}.
\end{equation}
Here $\alpha,\beta>0$. The residual consists of the imaginary mixed directions, giving
$W_{\rm F}=\sum_{\alpha,\beta>0}D_{\beta\alpha}^2/\ln2$.

For logical input $\rho$, tracing the Schr\"odinger commutator over
$MG_a$ gives the derivative at zero coupling,
\begin{equation}
 \begin{aligned}
 \dot\rho_B&=\frac{i}{g}\sum_\alpha\Tr(\rho A_\alpha)(C_\alpha-C_\alpha^\dagger)\\
 &=\frac1g\sum_{\beta>0} B_\beta
 \left(b_\beta+\sum_{\alpha>0}D_{\beta\alpha}\Tr(\rho A_\alpha)\right).
 \end{aligned}
\end{equation}
For the normalized Choi state this gives
\begin{equation}
 \dot\rho_{RB}=\frac1{dg}\sum_{\beta>0}
 \left(b_\beta I_R+\sum_{\alpha>0}D_{\beta\alpha}A_\alpha^{\mathsf T}
 \right)\otimes B_\beta .
 \label{eq:dimensional-choi-derivative}
\end{equation}
At zero coupling $\rho_{RB}=I/(dg)$, $\rho_B=I/g$, and
$\rho_R=I/d$ at all couplings. Around $I/n$, the entropy in bits has
quadratic coefficient $-n\,\Tr(\dot\rho^2)/(2\ln2)$ along a
trace-preserving path. Thus the coefficient in $I(R:B)$ is
$[dg\,\Tr(\dot\rho_{RB}^2)-g\,\Tr(\dot\rho_B^2)]/(2\ln2)$, where
$\dot\rho_B$ here uses $\rho=I/d$. Trace orthogonality cancels the bias
and leaves $\|D\|_{\rm F}^2/(2\ln 2)$, proving
Eq.~\eqref{eq:dimensional-leakage}. Purity of the full Choi vector then
gives the exact identity $I_c(R\rangle A)=\log_2 d-I(R:B)$ before decoding.

For a pure input $\ket\psi$, put
$L_\beta=b_\beta I+\sum_{\alpha>0}D_{\beta\alpha}A_\alpha$ and
$T=\sum_{\beta>0}L_\beta\otimes L_\beta$. Since $S(A)=S(B)$ and
$\Tr(B_\beta B_\nu)=g\delta_{\beta\nu}$, the quadratic coefficient of
$\ln g-S(A)$ in nats is
$\tfrac g2\Tr(\dot\rho_B^2)=\tfrac12\sum_{\beta>0}
\bigl(\bra\psi L_\beta\ket\psi\bigr)^2$. It equals $c/2$ for every pure input
if and only if
\begin{equation}
 P_{\rm sym}T P_{\rm sym}=cP_{\rm sym},
 \qquad P_{\rm sym}=\tfrac12(I+\mathrm{Swap}) .
 \label{eq:dimensional-flatness}
\end{equation}
Indeed, twice the coefficient is the expectation of $T$ on
$\ket\psi^{\otimes2}$. Constancy implies by homogeneity that
$\bra{z}^{\otimes2}(T-cI)\ket{z}^{\otimes2}=0$ for every complex $z$.
The coefficients of $\bar z_i\bar z_j z_k z_l$ for $i\le j$, $k\le l$
are the symmetric-subspace matrix elements up to nonzero normalization
factors, so each vanishes. The converse follows by taking the same
expectation in Eq.~\eqref{eq:dimensional-flatness}.

\section{The stabilizer curvature and the controlled gate}\label{app:quadratic}

Let $s_P=\langle J_0|P|J_0\rangle$ and $c_P(\epsilon)=\langle J_\epsilon|P|J_\epsilon\rangle$. For the Paulis with $s_P=\pm1$, $c'_P(0)=0$ because $J_0$ is a $P$ eigenstate. All other Paulis enter $\sum c_P^4$ only at fourth order. Differentiating the surviving terms twice and using the stabilizer projector $2^{-n}\sum_{P:s_P\ne0}s_PP=|J_0\rangle\langle J_0|$ gives
\begin{equation}
 \begin{aligned}
 &\left.\frac{d^2}{d\epsilon^2}2^{-n}\sum_P c_P^4\right|_0\\
 &\quad=-4\,2^{-n}\!\sum_{P:s_P\ne0} s_P\langle [H,[H,P]]\rangle
 =-8\Var_{J_0}(H).
 \end{aligned}
\end{equation}
The fourth moment starts at one, so its negative base-two logarithm has quadratic coefficient $4\Var(H)/\ln2$. This derivation uses only the stabilizer projector and applies to any pure qubit stabilizer base.

For the controlled rotation,
\begin{equation}
 H=\tfrac14(X_b-Z_M X_b-Z_a X_b+Z_M Z_a X_b).
\end{equation}
The first three terms lie in $\mathcal K_\partial$. The last Pauli sends the Bell product into the imaginary quadrature orthogonal to that real span, with unit norm before the factor $1/4$ (Eq.~\eqref{eq:mechanism-witness} and its proof). Thus the minimum squared norm is $1/16$. Among the 27 bare products of three nonidentity Paulis, the nine with equal bond axes fix the bond up to a sign and give zero residual. Each of the other eighteen has squared residual one. Six of those eighteen have matter factor $Y_M$, whose expectation vanishes everywhere on the real meridian. Testing only those logical inputs would miss their directional dependence.

\section{Two bonds and two candidate cuts}\label{app:two-bond}

Let $\phi(t)=\sqrt{t}\ket{00}+\sqrt{1-t}\ket{11}$ and prepare $\ket{\Phi^+}_{RM}\phi(1/2)_{a_L b_L}\phi(3/5)_{a_R b_R}$. The six-qubit Pauli fourth moment is $601/625$, giving base $\Mtwo=\log_2(625/601)$. The right bond already carries magic, so we use an operator norm for a separate support calculation. Expand $H_2=P_{1,M}P_{1,a_L}X_{b_L}X_{a_R}X_{b_R}$ into four orthogonal Pauli strings of equal absolute coefficient. Two of them have support on all three blocks of $RM|a_Lb_L|a_Rb_R$. The relative Frobenius norm left after projecting off all pairwise strings is therefore $1/\sqrt2$. For $R|M|G_{\rm all}$ the operator is supported wholly on $MG_{\rm all}$ and the residual is exactly zero.

After applying $e^{i\pi H_2/4}$ and conditioning on $\psi_q$, take the smaller of $S(a_L)$ and $S(a_R)$ at each of 401 equally spaced values of $q\in[0,1]$. The right leg has smaller entropy at $q=0$. The first sampled point at which the left leg has smaller entropy is $q=0.335$, and the range of the sampled minimum is $0.248909$ nats. These are entropies of two different legs. To interpret a switch as a change of extremal surface for one boundary region would require a common reconstruction prescription for that region.

\section{Spin-chain information and numerical checks}\label{app:spin-checks}

For the cyclic injection $O=\operatorname{diag}(-1,1,1)$ and the proper rotation $-O$ can be removed by a qubit unitary. The local Choi state is therefore equivalent to that of a depolarizing channel with signed parameter $x=-C_{j0}s(\epsilon)$. Its eigenvalues are $(1+3x)/4$ and $(1-x)/4$ three times. Both marginals are maximally mixed, so the exact mutual information is
\[
 f(x)=\frac{(1+3x)\ln(1+3x)+3(1-x)\ln(1-x)}{4\ln2}.
\]
Here $f'(0)=0$, $f''(0)=3/\ln2$, and $s'(0)=2$. These derivatives reproduce Eq.~\eqref{eq:transported-curvature}. The signed argument matters away from the origin. Spin isotropy also matters. For $U=\exp(-i\pi Z_0Z_1/4)$, the local $Z$ polarization stays at site 0 while the $X,Y$ components spread onto both spins. One of the three Pauli components survives locally and $W_0/\Wpartial=1/3$, although the longitudinal transfer is one.

For a set $S$ of spins, define $W_S=\partial_\epsilon^2 I(R:S)|_0$. Write $c_{\alpha P}=2^{-N}\Tr(P U_t\sigma_{\alpha,0} U_t^\dagger)$ for $\alpha=x,y,z$. Tracing out the spins outside $S$ keeps exactly the Pauli strings with support in $S$. At zero coupling the state on $RS$ is maximally mixed, so the quadratic entropy expansion gives
\begin{equation}
 \frac{W_S}{\Wpartial}=\frac13\sum_{\alpha=x,y,z}
    \sum_{\operatorname{supp}P\subseteq S}|c_{\alpha P}|^2 .
 \label{eq:spin-subset}
\end{equation}
The factor $1/3$ comes from the equal strengths of the three injected polarizations. To obtain the bins in Fig.~\ref{fig:spin-chain}, group each string once by the size of its support. A string contained in two overlapping regions contributes to the curvature of each region.

For the coherent two-spin control write $U=(I-i\,\mathrm{SWAP})/\sqrt 2$. In the evolved operator $U(Z\otimes I)U^\dagger$, $Z\otimes I$ and $I\otimes Z$ each have coefficient $1/2$. The commutator with SWAP produces the two-spin remainder with squared weight $1/2$. The random channel gives just the average of the two one-spin operators. Applying the normalized Choi entropy expansion of Eq.~\eqref{eq:leakage} to these Pauli coefficients gives the two joint curvatures. Directly tracing the normalized reference--two-spin state reproduced all four single-site predictions and both joint predictions to within $3.9\times10^{-9}$ bits per unit $\epsilon^2$.

For the large chain we represented $U_t Z_0 U_t^\dagger$ as a matrix-product operator in the normalized Pauli basis. Each symmetric time step applies exact nearest-neighbor exchange gates with half steps in ascending and then descending bond order. No renormalization was applied. Figure~\ref{fig:spin-chain} uses step $J\,\Delta t/\hbar=1/32$, singular-value cutoff $10^{-12}$ and maximum bond dimensions 72 for the signed profiles and 96 for the support fractions. At bond dimension 48, halving the step from $1/16$ to $1/32$ changed any sampled single-site coefficient by at most $6.2\times10^{-5}$. Increasing the bond dimension from 72 to 96 changed any reported support fraction by less than $1.5\times10^{-5}$. At bond dimension 96, halving the time step from $1/32$ to $1/64$ changed each sampled support fraction by less than $1.1\times10^{-5}$. The largest signed total-spin drift in the original three resolution runs was below $9.5\times10^{-5}$.

An independent calculation exponentiated the full chain Hamiltonian on six spins. At the tested times through $2\hbar/J$, it reproduced the matrix-product one-site coefficients to within $1.6\times10^{-5}$. On four bath spins we prepared the reference and encoder explicitly, traced away the retained code qubits, evolved the bath, and formed the marginal entropies. Symmetric differences at $\epsilon=\pm0.0005$ gave the predicted site curvatures to within $1.1\times10^{-5}$ bits per unit $\epsilon^2$. Enumeration of all 256 four-site Pauli products independently verified the support-bin algorithm. The small-chain checks use the full density matrices. The 41-spin calculation uses the analytical relation to the spin correlator.

Separate Lean~4 proofs verify the scalar form forced by two quarter-turn symmetries, the partial-SWAP marginals, and the half-weight two-spin remainder at square-root SWAP. They also verify the curvature of the Bell-information function under a smooth injection and the conservation and projection identities given their hypotheses. The physical identification of the Choi spectrum with the four probabilities above is analytic. The fractions in Fig.~\ref{fig:spin-chain} were obtained by numerical evolution.

\section{Pairwise operations on reference, matter and geometry}
\label{sec:resource-rank}

Grouping the two bond legs into one geometry block changes the removable
directions. Indeed, every physical generator on $MG_aG_b$ is supported on
the two blocks $M$ and $G=G_aG_b$.
The stronger question is whether pairwise generators can reproduce the
action of an arbitrary generator that also acts on $R$. A Schmidt-rank
count answers this question for a class of finite-dimensional bases.

\begin{proposition}[Pairwise tangent dimension]
\label{prop:resource-rank}
Suppose the base is $\ket{\chi}_{RM}\ket{0}_G$, where $R,M,G$ have
dimensions $d_R,d_M,g$ and $\ket{\chi}$ has Schmidt rank $k$. Allow every
Hermitian on any pair of the three blocks. The real dimension of their
centered actions is
\begin{equation}
  2(d_Rd_M-1)
  +2(g-1)\big[d_Rd_M-(d_R-k)(d_M-k)\big].
  \label{eq:resource-rank}
\end{equation}
For $g>1$ this dimension equals the full projective dimension if and only if $k=\min(d_R,d_M)$.
\end{proposition}

\begin{proof}
Let $S_R,S_M$ be the Schmidt supports of $\ket{\chi}$.
In the sector with geometry state $\ket{0}$, an arbitrary Hermitian on
$RM$ produces every centered tangent: for $v\perp\chi$ the operator
$\ket v\bra\chi+\ket\chi\bra v$ sends $\ket\chi$ to $v$. This gives the
first term of \eqref{eq:resource-rank}.

Fix a basis vector $\ket{\mu}\perp\ket{0}$ of $G$. On $RG$ we can use
the Hermitian operator
$X_R\otimes\ket{\mu}\bra0+X_R^\dagger\otimes\ket0\bra{\mu}$ for any
complex $X_R$. It produces $X_R\ket{\chi}\otimes\ket{\mu}$.
Replacing $X_R$ by $iX_R$ realizes the other real quadrature. Varying
$X_R$ gives $\mathcal H_R\otimes S_M$ in this sector because the $k$
nonzero Schmidt coefficients allow the $k$ corresponding columns of
$X_R$ to be chosen independently. The $MG$ generators similarly give
$S_R\otimes\mathcal H_M$. The complex orthogonal complement of their sum
is $S_R^\perp\otimes S_M^\perp$, of dimension $(d_R-k)(d_M-k)$.
An $RM$ generator has zero component in the $\mu$ sector. Summing the real
dimensions over the $g-1$ orthogonal sectors proves the formula.
\end{proof}

For the reference-inclusive partition of the four-qubit Choi state,
$(d_R,d_M,g,k)=(2,2,4,2)$ gives dimension $30$, the full projective
dimension. Enlarging the pure geometry block preserves this conclusion
for any full-Schmidt-rank reference--matter state. If both Schmidt
supports are proper and $g>1$, any product of vectors from
$S_R^\perp$, $S_M^\perp$ and the orthogonal complement of $\ket{0}_G$ provides a missed tangent.
Equation~\eqref{eq:resource-rank} concerns centered state vectors at
these more general bases. The identification with a stabilizer R\'enyi
coefficient uses the additional stabilizer-base hypothesis.

\bibliography{references}
\end{document}